\documentclass[sigconf]{acmart}

\AtBeginDocument{%
  \providecommand\BibTeX{{%
    \normalfont B\kern-0.5em{\scshape i\kern-0.25em b}\kern-0.8em\TeX}}}
    
\copyrightyear{2021} 
\acmYear{2021} 
\setcopyright{acmcopyright}\acmConference[SIGIR '21]{Proceedings of the 44th International ACM SIGIR Conference on Research and Development in Information Retrieval}{July 11--15, 2021}{Virtual Event, Canada}
\acmBooktitle{Proceedings of the 44th International ACM SIGIR Conference on Research and Development in Information Retrieval (SIGIR '21), July 11--15, 2021, Virtual Event, Canada}
\acmPrice{15.00}
\acmDOI{10.1145/3404835.3462988}
\acmISBN{978-1-4503-8037-9/21/07}

\usepackage{times}
\usepackage[utf8]{inputenc}
\usepackage{graphicx}
\usepackage{amsmath}
\usepackage{algorithm}
\usepackage{algorithmic}
\usepackage{textcomp}
\usepackage{CJKutf8}
\usepackage{subcaption}
\usepackage{balance}
\usepackage{multirow}
\usepackage[skip=3.5pt]{caption}
\usepackage[compact]{titlesec}

\newcommand{\pqm}{Poeem}

\newcommand{\ie}{\textit{i.e.}}

\newcommand{\R}{\mathbb{R}}

\newcommand{\x}{\mathbf{x}}
\newcommand{\y}{\mathbf{y}}
\renewcommand{\v}{\mathbf{v}}
\renewcommand{\c}{\mathbf{c}}
\newcommand{\CC}{\mathcal{C}}
\newcommand{\T}{\mathcal{T}}
\renewcommand{\H}{\mathcal{H}}
\newcommand{\sg}{\textrm{sg}}

\newcommand\blfootnote[1]{%
  \begingroup
  \renewcommand\thefootnote{}\footnote{#1}%
  \addtocounter{footnote}{-1}%
  \endgroup
}

\begin{document}
\fancyhead{}

\setlength{\abovedisplayskip}{2.5pt}
\setlength{\belowdisplayskip}{3pt}

\title{Joint Learning of Deep Retrieval Model and Product Quantization based Embedding Index}

\author{Han Zhang$\,^{1}$, Hongwei Shen$\,^{2}$, Yiming Qiu$\,^{1}$, Yunjiang Jiang$\,^{2}$, Songlin Wang$\,^{1}$, Sulong Xu$\,^{1}$, \\ Yun Xiao$\,^{2}$, Bo Long$\,^{1}$ and Wen-Yun Yang$\,^{2*}$}
\affiliation{%
  \institution{$^{1}\,$JD.com, Beijing, China; $^{2}\,$JD.com Silicon Valley Research Center, Mountain View, CA, United States}
  \{\small zhanghan33, hongwei.shen1, qiuyiming3, yunjiang.jiang, wangsonglin3, xusulong, xiaoyun1, bo.long, wenyun.yang\}@jd.com
}


\begin{abstract}
Embedding index that enables fast approximate nearest neighbor (ANN) search, serves as an indispensable component for state-of-the-art deep retrieval systems. Traditional approaches, often separating the two steps of embedding learning and index building, incur additional indexing time and decayed retrieval accuracy.
In this paper, we propose a novel method called \emph{{\pqm}}, which stands for \textbf{p}roduct quantizati\textbf{o}n based \textbf{e}mbedding index jointly trained with deep r\textbf{e}trieval \textbf{m}odel, to unify the two separate steps within an end-to-end training, by utilizing a few techniques including the gradient straight-through estimator, warm start strategy, optimal space decomposition and Givens rotation.
Extensive experimental results show that the proposed method not only improves retrieval accuracy significantly but also reduces the indexing time to almost none. 
We have open sourced our approach\footnote{https://github.com/jdcomsearch/poeem} for the sake of comparison and reproducibility. 
\blfootnote{$^*\,$ Corresponding author}
\end{abstract}


\maketitle

\section{Introduction}
\label{sec:introduction}

% introduce traditional indexes for search (inverted index), recommendation (key value store) and ads (tree based index)
Various types of indexes play an indispensable role in modern computational systems to enable fast information retrieval. As a traditional one,
inverted index~\cite{dean2009challenges} has been widely used in web search,  e-commerce search, recommendation and advertising in the past few decades. Recently, with the advent of the deep learning era, embedding indexes~\cite{huang2013learning,youtube2016}, which embed user/query and item in a latent vector space, show excellent performance in many industrial retrieval systems~\cite{huang2020embedding,zhang2020towards,mind2019}. 
Embedding index enjoys several appealing advantages: a) the embeddings can be learned to optimize downstream retrieval task of interests, and b) efficient algorithms for maximum inner product
search (MIPS) or approximate nearest neighbors (ANN), such as LSH~\cite{datar2004locality}, Annoy~\cite{Github:annoy} and state-of-the-art product quantization (PQ) based approaches~\cite{jegou2010product,johnson2019billion,guo2020accelerating}, can be leveraged to retrieve items in a few milliseconds.
%Along with the open-sourced libraries Faiss~\cite{johnson2019billion} and ScaNN~\cite{guo2020accelerating}, PQ based embedding indexes are widely adopted in many industrial systems~\cite{huang2020embedding,zhang2020towards}.

% talk about drawbacks of separately built index, and introduce joint training of tree based index, and others. But they are cumbersome, hard to implement, and too many tricks
Embedding indexes, however, also suffer from a few drawbacks. The major one resides in the separation between model training and index building, which results in extra index building time and decayed retrieval accuracy. 
Thus, recently, there is a new trend of abandoning separately built embedding indexes but embracing jointly learned structural indexes which have shown improved performance than the former. In general, the approaches with joint learning structure can be summarized into two types, tree based ones~\cite{zhu2018learning,tdm2} and PQ based ones~\cite{yu2018product,cao2016deep,klein2017defense}. Those tree based approaches normally require special approximate training techniques, whose complications slow down their wide adoptions. 
Those existing PQ based approaches are designed for only small computer vision tasks, such as retrieval from tens of thousands of images, thus inapplicable to large scale information retrieval tasks with at least millions of items, such as what we have in a real-world industrial retrieval system.
% The existed product quantization approaches dont't have any quantizer skills, like coarse quantization~\cite{jegou2010product}, to improve retrieval efficiency, thus are inapplicable to large scale industrial task.
% Most product quantization based approaches have suboptimal structures which can be improved in some aspects, such as ~\cite{cao2016deep} introduces a deep quantization network which update codewords by k-means whereas ignore the label information, ~\cite{yu2018product} proposes a soft product quantization network which is not optimal in retrieval efficiency.

In this paper, we advance the approach of product quantization based embedding index jointly trained with deep retrieval model.
It is not trivial and we have to overcome a few hurdles by appropriate techniques: 
1) the quantization steps, as the core of PQ based embedding indexes, have non-differentiable operations, such as $\arg\min$, which disable the standard back propagation training. Thus, we leverage the gradient straight-through estimator~\cite{bengio2013estimating} to bypass the non-differentiability in order to achieve end-to-end training.
2) The randomly initialized quantization centroids lead to very sparse centroid assignments, low parameter utilization and consequentially higher quantization distortion. Thus, we introduce a warm start strategy to achieve more uniform distributed centroid assigments.
3) The standard optimized product quantization (OPQ)~\cite{ge2013optimized} algorithm, which rotates the space by an orthonormal matrix to further reduce PQ distortion, can not iteratively run together with the joint model training. Thus, we develop a steepest block coordinate descent algorithm with Givens rotations~\cite{matrix_computations} to learn the orthonormal matrix in this end-to-end training. 

% However, the quantization steps, as the core of PQ based embedding indexes, have non-differentiable operations, which make the joint learning of embedding index a non-trivial problem.
% We carefully design the algorithm to make the joint training possible and optimal, by a few techniques: the straight-through estimator~\cite{bengio2013estimating} that overcomes the non-differentiability in  quantization steps to enable end-to-end training, a warm start strategy of quantization centroids that reduces centroid assignment sparsity in order to reduce quantization distortion, and a Givens rotation based block coordinate descent algorithm that iteratively learns an orthonormal matrix to allow PQ operated with optimized space decomposition. 

As a result, our proposed method \emph{{\pqm}}, which stands for \textbf{p}roduct quantizati\textbf{o}n based \textbf{e}mbedding index jointly trained with deep r\textbf{e}trieval \textbf{m}odel, 
enjoys advantages of almost no index building time and no decayed retrieval accuracy. Aiming at a more practical approach ready for wide adoptions, our method is capsulated in a standalone indexing layer, which can be easily plugged into any embedding retrieval models.
\section{Method}
\label{sec:method}
% In this section we present our method step by step as follows: in Section~\ref{sec:retrieval_model} we introduce the basic architecture of standard retrieval model, in Section~\ref{sec:indexing_layer} we introduce an indexing layer that could be easily plugged in any retrieval model to produce ready-to-user embedding index right after training, in Section~\ref{sec:givens_rotation}, we introduce a novel algorithm to learn an rotation matrix together to further reduce quantization distortion, in Section~\ref{sec:algorithm} we present the full algorithm of training the indexing layers and rotation matrix, and in Section~\ref{sec:serving}, we talk about the post training steps for fast online retrieval.

\subsection{Revisiting Retrieval Model}
\label{sec:retrieval_model}
\begin{figure*}[t]
    \centering
    \includegraphics[width=\textwidth]{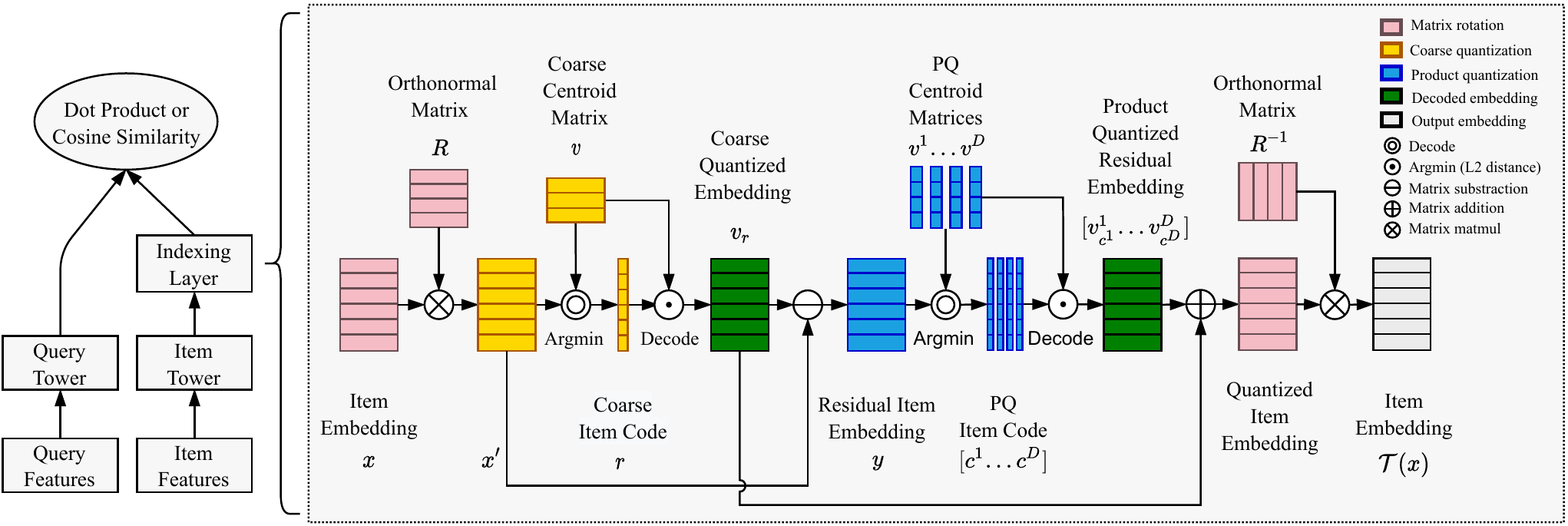}
    \caption{Illustration of a retrieval model with an embedding indexing layer, which is composed of four steps, coarse quantization function $\psi$ (yellow), product quantization function $\phi$ (blue), decoder function $\rho$ (green), and givens rotation (pink). }
    \label{fig:overview}
    \vspace{-0.1in}
\end{figure*}

A standard embedding retrieval model, as shown at the left side of Figure~\ref{fig:overview}, is composed of a query tower $Q$ and an item tower $S$. Thus, for a given query $q$ and an item $s$, the scoring output of the model is
\begin{equation}
f(q, s) = F(Q(q), S(s)) \label{eq:scoring}
\end{equation}
where $Q(q) \in \R^{d}$ denotes query tower output embeddings in $d$-dimensional space. Similarly, $S(s) \in \R^{d}$ denotes an item tower output in the same dimensional space. The scoring function $F(.,.)$, usually inner product or cosine value, computes the final score between the query and item, and has been proven to be successful in many applications~\cite{youtube2016,huang2020embedding,zhang2020towards}. 
After the model is trained, traditional approaches still require an additional step -- computing item embeddings and building an embedding index for them -- before it is ready for online serving. 
% Faiss~\cite{johnson2019billion} and ScaNN~\cite{guo2020accelerating}, based on product quantization~\cite{jegou2010product}, are two typical embedding index libraries widely used in industry. 
However, this additional step not only spends extra time to build index, but also causes decay of recall rate~\cite{jegou2010product}.
In the following sections, we will present our approach to the rescue of these two shortcomings.

\subsection{Embedding Indexing Layer}
\label{sec:indexing_layer}

Figure~\ref{fig:overview} illustrates an overview of the proposed embedding indexing layer inside a typical deep retrieval model. Formally, the indexing layer defines a \emph{full quantization function} $\T: \R^d \xrightarrow{}\R^d$ that maps an input embedding $\x$ to an output embedding $\T(\x)$, which can be decomposed into four functions: a \emph{coarse quantization} function $\psi$, a \emph{product quantization} function $\phi$ and a \emph{decoder} function $\rho$, and a \emph{rotation} function with an orthonormal matrix $R$.
%the indexing layer also multiplies the input embedding $\x$ by a \emph{rotation matrix} $R$ and the output embedding $\T(\x)$ by its inverse matrix $R^{-1}$. 
Note that we are using orthonormal matrix and rotation matrix interchangeably. Now let's explain in detail these operations in the following sections.

% 1) a \emph{coarse quantization} function $\psi: \R^d \xrightarrow{} \{1,\cdots,J\}$, that maps a continuous vector into a $J$-way discrete code, 2) a \emph{product quantization} function $\phi: \R^d \xrightarrow{} \{1,\cdots,K\}^D$, that maps a continuous vector into a $K$-way $D$-dimensional discrete code, and 3) a \emph{decoder} function $\rho: \left(\{1,\cdots,J\}, \{1,\cdots,K\}^D\right) \xrightarrow{} \R^d$ that maps the discrete code back to a continuous vector. Thus, the function $\T$ can be expressed as $\T(\cdot) = \rho \circ \phi(\cdot)$.

The \textbf{coarse quantization} function $\psi: \R^d \xrightarrow{} \{1,\cdots,J\}$, which maps a continuous vector $\x$ into a $J$-way discrete \emph{coarse code} $r$ by a coarse centroid matrix $\v \in \R^{J \times d}$, can be defined as follows
\begin{equation}
    \psi(\x) = r = \arg \min_k \textrm{dist}(\x, \v_k)
    \label{eq:coarse_argmin}
\end{equation}
where $\textrm{dist}(.,.)$ stands for a distance measure, typically L2 distance, between two vectors, and the vector $\v_k \in \R^d$ stands for the $k$-th centroid, \ie, $k$-th row in $\v$. It is not hard to see that this function just finds the nearest centroid for a given input vector $\x$, and outputs the centroid index. Thus, a coarse quantization residual exists, formally as follows,
\[
\y = \x - \v_r.
\] 
We will further deal with it by product quantization.

% \subsubsection{Product Quantization}
The \textbf{product quantization (PQ)} function $\phi: \R^d \xrightarrow{} \{1,\cdots,K\}^D$, which maps the above residual vector $\y$ into a $K$-way $D$-dimensional \emph{PQ code} $\c$, can be defined as follows.
First, we denote the vector $\y$ as concatenation of $D$ subvectors: 
\begin{equation}
    \y=[\y^1, \ldots, \y^D]. \label{eq:decomp}
\end{equation}
For simplicity, it is a common practice to divide the original dimension $d$ evenly. In other words, each subvector is $\y^j \in \R^{d/D}$. The Cartesian product $\CC = \CC^1 \times \ldots \times \CC^{D}$ is the set in which a \emph{PQ code} $\c\in \CC$ is formed by concatenating the $D$ sub-codewords: $\c=[c^1, \ldots, c^D ]$, with each $c^j \in \CC^j$. Note that the $D$ codes can be computed independently as follows
\begin{equation}
    c^j = \arg \min_k \textrm{dist}(\y^j, \v_k^j)
    \label{eq:argmin}
\end{equation}
where $\v_k^j$ stands for the $k$-th centroid for the $j$-th subvectors. Finally, we can define the function $\phi$ as follows
\begin{equation} 
\phi(\y) = [c^1, \ldots, c^D].
\label{eq:pqcode}
\end{equation}

% \subsubsection{Decoder Function}
The \textbf{decoder} function 
\[\rho: \left(\{1,\cdots,J\}, \{1,\cdots,K\}^D\right) \xrightarrow{} \R^d\]
that maps the discrete codes back to a continuous vector, can be formally defined as follows,
\begin{equation*}
    \rho(r, [c^1, \ldots, c^D]) = \v_r + [\v^1_{c^{1}}, \ldots, \v^D_{c^{D}}]
\end{equation*}
which sums over the coarse quantized vector $\v_c$, and the concatenation of product quantized vector $\v^j_{c^j}$.

% \subsubsection{Quantization Function}

% \subsubsection{Rotation Matrix}
The \textbf{rotation} function by an orthonormal matrix $R$, also known as optimized product quantization (OPQ)~\cite{ge2013optimized}, rotates the input embedding $\x$ by
\begin{align*}
    \x' &= R \x 
\end{align*}
to further reduce the quantization distortion. Since the rotation allows product quantization to operate in a transformed space, thus relaxes the constraints on PQ centroids. For example, note that any reordering of the dimensions can be represented by a rotation matrix. 

After the above decoder function, we can also ``rotate back'' to the original space by $R^{-1}$, so that this rotation is fully encapsulated inside the standalone indexing layer. Also note that the inverse of an orthonormal matrix is just its transpose~\cite{matrix_computations}, \ie, $R^{-1} = R^\top$, which is very cheap to compute.

Finally, with the above four functions, we can now define the \textbf{full quantization} function $\T$ as follows
\begin{equation}
    \T(\x) = R^\top \rho (\psi(\x'), \phi(\x' - \v_{\psi(\x')}))
    \label{eq:tau}
\end{equation}
where $\x'=R\x$.

\subsection{Training Algorithm}
\label{sec:algorithm}
\subsubsection{Loss Function}
Straightforward optimization of the above embedding indexing layer by the standard back propagation algorithm is infeasible, as the $\arg\min$ operation in Eqs. (\ref{eq:coarse_argmin}) and (\ref{eq:argmin}) is non-differentiable. 
In fact, this is the difficulty that prevents previous researchers from training embedding indexes jointly with retrieval model. 
Here we propose to leverage the gradient straight-through estimator~\cite{bengio2013estimating} by adjusting the original quantization function $\T$ in Eq. (\ref{eq:tau}) as follows
\begin{equation}
    \H (\x) = \x - \sg(\x - \T(\x)) 
    \label{eq:approx_t}
\end{equation}
where $\sg$ is the \emph{stop gradient} operation. 
During the forward pass, the quantized embedding $\T(\x)$ is emitted; During the backward pass, the gradient is passed to the original embedding $\x$ directly, bypassing the entire quantization function $\T$. Note that Eq. (\ref{eq:approx_t}) only approximates gradient for original embedding $\x$, does not update the quantization parameters (centroids) in the back propagation. Similar as previous works~\cite{van2017neural,chen2020differentiable}, we add a regularization term into the loss function to minimize the quantization distortion as follows,
\begin{equation*}
    \mathcal{L}_{reg} = \sum_\x||\T(\x) - \sg(\x)||^2
\end{equation*}
which essentially updates the coarse and PQ centroids to be arithmetic mean of their members.

\subsubsection{Warm Start Centroids}
In practice, we find that random initialization of the quantization parameters (the centroids $\v_k$ and $\v_k^j$ in Eqs. (\ref{eq:coarse_argmin}) and (\ref{eq:argmin})) results in very sparse centroid assignments, which consequentially hurts the retrieval quality (see experiments, Section~\ref{sec:warm_start}). Fortunately, we are able to overcome this hurdle by simply warm starting the centroids. In detail, we train the model without the above indexing layer for a number of warm-start steps, then we plug in the indexing layer with centroids initialized by a standard k-means clustering. 
%More details can be found in full training algorithm shown in Algorithm~\ref{alg:train}.

\subsubsection{Givens Rotation}
\label{sec:givens_rotation}
% \vspace{-1.5mm}
% \setlength{\parskip}{2mm}

% To overcome the, we develop the following approach to learning this orthonormal matrix jointly within the indexing layer.
Learning the optimal rotation matrix with fixed embeddings $\x$, previously formulated as the Orthogonal Procrustes problem~\cite{ge2013optimized,procrustes1966}, is incompatible with our end-to-end training, since the embeddings $\x$ is not fixed. 
Owing to the previous work~\cite{hurwitz1963ueber} which shows that any rotation matrix can be represented by a product of Givens rotations~\cite{matrix_computations}, we are able to jointly learn the rotation matrix by a steepest block coordinate descent algorithm~\cite{wright2015coordinate, beck2013convergence} with each coordinate as a specific Givens rotation within two axes.

\begin{figure*}[t]
    \centering
    \begin{subfigure}[b]{0.3\textwidth}
        \centering
        \includegraphics[width=\textwidth]{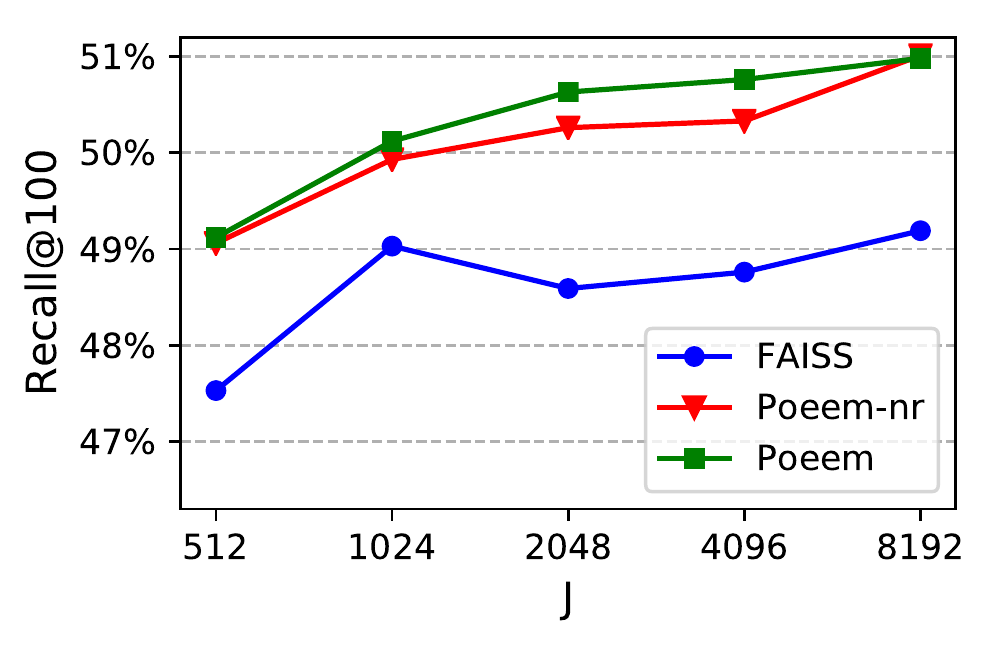}
        \vspace{-6mm}
        \caption{Parameter $J$}
        \label{fig:compare_j}
    \end{subfigure}
    \begin{subfigure}[b]{0.3\textwidth}
        \centering
        \includegraphics[width=\textwidth]{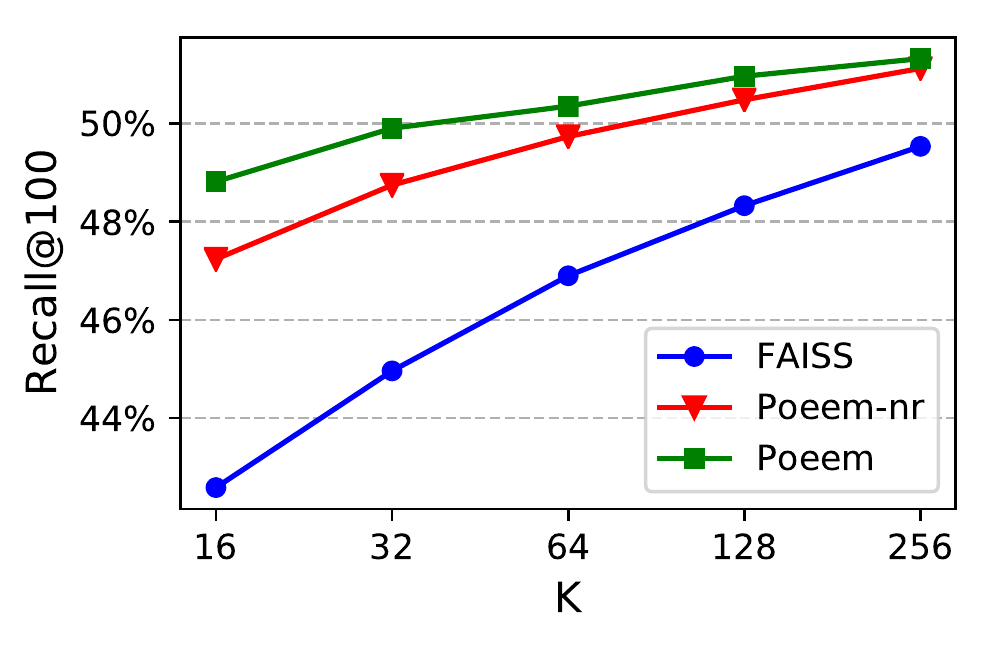}
        \vspace{-6mm}
        \caption{Parameter $K$}
        \label{fig:compare_K}
    \end{subfigure}
    \begin{subfigure}[b]{0.3\textwidth}
        \centering
        \includegraphics[width=\textwidth]{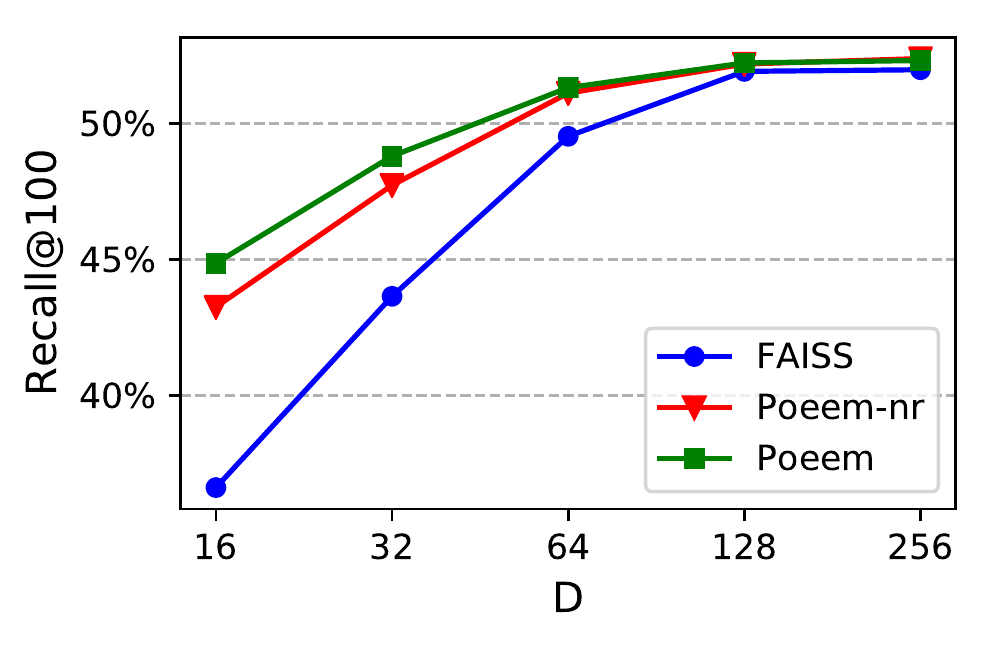}
        \vspace{-6mm}
        \caption{Parameter $D$}
        \label{fig:compare_D}
    \end{subfigure}
    \caption{Comparison between baseline (offline indexing with Faiss) and the proposed {\pqm} for different values of $J$, $K$ and $D$. \pqm-nr stands for the variant of {\pqm} without using rotation matrix.}
    \vspace{-2mm}
    \label{fig:comparison}
\end{figure*}

\section{Experiment}
\label{sec:experiment}
% In this section, we present experiment results and ablation studies to demonstrate the effectiveness of our proposed method {\pqm}. Specifically, in Section~\ref{sec:setup}, we explain the datasets, hyperparameter setup, evaluation metrics and training environments; In Section~\ref{sec:comparison}, we show the comparison results with traditional offline indexing approach; In Section~\ref{sec:distortion}, we study the relations between quantization distortion and recall rate, and the effects of three techniques to reduce the quantization distortion in order to improve the recall rate; in Section~\ref{sec:computational_cost}, we present the computational costs of the proposed method.

\subsection{Setup}
\label{sec:setup}
% talk about details of the two datasets, evaluation metrics, training setup, parameters, implementation in Tensorflow, and so on.

In Table~\ref{tab:dataset}, we evaluate our methods on three datasets: a JD.com search click log data where a user input query is used to retrieve items, and two public datasets of MovieLens~\cite{harper2015movielens} and Amazon Books~\cite{he2016ups} where user historical behavior is used to retrieve next behavior. 
We evaluate retrieval accuracy by precision@$k$ (p@$k$) and recall@$k$ (r@$k$) metrics, which are standard retrieval quality metrics. %In our experiments, we choose k=100.
% Specifically, for each query or user, we retrieve from embedding indexes a set of top k items, which is then compared with the ground truth set of items to calculate the precision and recall rates.

We implement {\pqm} in Tensorflow 1.15, and train models in a single machine with 4 Nvidia V100 GPU cards, and evaluate all methods in a 48-cores CPU machine. A typical set of parameters are as follows: a two-tower retrieval model with cosine scoring and hinge loss of margin $0.1$, embedding size $512$ for private dataset, $128$ for MovieLens and $128$ for Amazon Books, Adagrad optimizer with learning rate $0.01$, and batch size $1024$.

% \begin{table}[tb]
%     \centering
%     \caption{Dataset statistics.}
%     \label{tab:dataset}
%     \begin{tabular}{c|ccc}
%     \hline
%          & Private &  MovieLens & Amazon\\
%     \hline
%         no. of examples & 9989135 & 9780514 & \\
%         feature type & text & id & id \\
%         vocab size & 50411 & & \\
%     \hline
%     \end{tabular}
% \end{table}

\begin{table}[t]
\centering
\caption{Dataset statistics.}
\small
\begin{tabular}{c|ccc}
\hline
Dataset   & \# Examples &  \# Users & \# Items  \\ \hline
Private   & 9,989,135   & 1,031,583 & 1,541,673  \\ 
MovieLens & 9,939,873   & 129,797  & 20,709    \\ 
Amazon Books    & 8,654,619   & 294,739  & 1,477,922 \\ \hline
\end{tabular}
\vspace{-4mm}
\label{tab:dataset}
\end{table}

\subsection{Comparison with Offline Indexing}
\label{sec:comparison}
Table~\ref{tab:comparison} presents comparison results with the baseline methods, offline indexing with LSH~\cite{datar2004locality}, Annoy~\cite{Github:annoy}, ScaNN~\cite{guo2020accelerating} and Faiss~\cite{johnson2019billion}.
To conduct fair comparisons, Faiss uses the index type of IVFPQ which shares the same product quantization parameters as {\pqm}. Since other baselines do not have similar product quantization structure, we choose parameters which have the same retrieval time cost as {\pqm}. %Specifically, it costs 60 seconds to retrieve top 100 items for 2 million queries of the private dataset, 0.2 second to retrieve top 100 items for 1000 users of MovieLens, and 7.6 seconds to retrieve top 100 items for 5000 users of Amazon Books. 
We can observe that the proposed {\pqm} outperforms all the baseline methods by precision@100 and recall@100.

\begin{table}[tb]
    \centering
    \caption{Comparison between the baseline methods and {\pqm}.}
    \label{tab:comparison}
    \small
    \setlength{\tabcolsep}{1mm}
    \begin{tabular}{c|cc|cc|cc}
    \hline
    \multirow{2}{*}{Method} & \multicolumn{2}{c|}{Private} & \multicolumn{2}{c|}{MovieLens} & \multicolumn{2}{c}{Amazon Books} \\ \cline{2-7} 
                            & p@100         & r@100        & p@100          & r@100         & p@100        & r@100        \\ \hline
    LSH & $1.28\%$ & $25.64\%$ & $7.48\%$ & $34.50\%$ & $0.51\%$ & $4.11\%$ \\
    Annoy & $1.24\%$ & $23.42\%$ & $7.85\%$ & $35.53\%$ & $0.72\%$ & $5.71\%$ \\
    ScaNN & $2.25\%$ & $47.30\%$ & $7.91\%$ & $36.50\%$ & $0.72\%$ & $5.71\%$ \\
    Faiss & $2.37\%$ & $49.54\%$ & $8.02\%$ & $36.72\%$ & $0.68\%$ & $5.46\%$ \\
    \hline
    \multirow{2}{*}{Poeem} & $\mathbf{2.42\%}$ & $\mathbf{51.13\%}$ & $\mathbf{8.22\%}$ & $\mathbf{37.48\%}$ & $\mathbf{0.73\%}$ & $\mathbf{5.90\%}$ \\
                         & $(+0.05\%)$ & $(+1.59\%)$ & $(+0.20\%)$ & $(+0.76\%)$ & $(+0.05\%)$ & $(+0.44\%)$ \\
    %Poeem-rotation & $\mathbf{2.43\%}$ & $\mathbf{51.34\%}$ & $\mathbf{8.08\%}$ & $\mathbf{36.75\%}$ & $\mathbf{0.72\%}$ & $\mathbf{5.64\%}$ \\ 
    \hline
    \end{tabular}
    \vspace{-4mm}
\end{table}

Figure~\ref{fig:comparison} shows the recall@$100$'s trend with varying values of $J$, $K$ and $D$, with or without $R$. It is clear that {\pqm} outperforms the baseline for all different parameter values. In detail, firstly, the rotation matrix has a positive impact on recall@$100$. Secondly, we can see a trend that, with smaller parameters, the gap between the proposed method and baseline is enlarged. This indicates that the best scenario of the proposed method is for indexes of large dataset where large compression ratio is required. Thirdly, the performance gap almost diminishes when $D$ approaches to $d$ ($512$ in this case), which, however, is actually impractical but just for reference purpose. Since if $D=d$, product quantization reduces to numerical quantization whose low compression ratio is infeasible for large data. 
Note that a proper choice of $J$ is necessary to prevent imbalance distribution of coarse codes and consequentially unstable retrieval time.

\iffalse
\subsection{Effects of Reducing Quantization Distortion}
\label{sec:distortion}
Figure~\ref{fig:coarse} illustrates the effects on quantization distortion and recall@$100$, for varying indexing layer parameters: $J$, $K$ and $D$.
The curves clearly show the same trend: lower quantization distortion leads to higher retrieval quality that measured by recall@$100$. Moreover, we can observe that the parameter $D$ has the largest effect on quantization distortion, though one needs to be aware that larger $D$ is also more expensive in space than the other two parameters. Since the $D$ dimensional item PQ codes compose the major part of embedding indexes. In contrast, the coarse quantization is much cheaper since it adds only one coarse code to each item. Thus, the curve of $J$ shows the retrieval quality gain with marginal space cost of coarse quantization.
\fi

\begin{figure}
    \centering
    \begin{subfigure}[b]{0.153\textwidth}
        \centering
        \includegraphics[width=\textwidth]{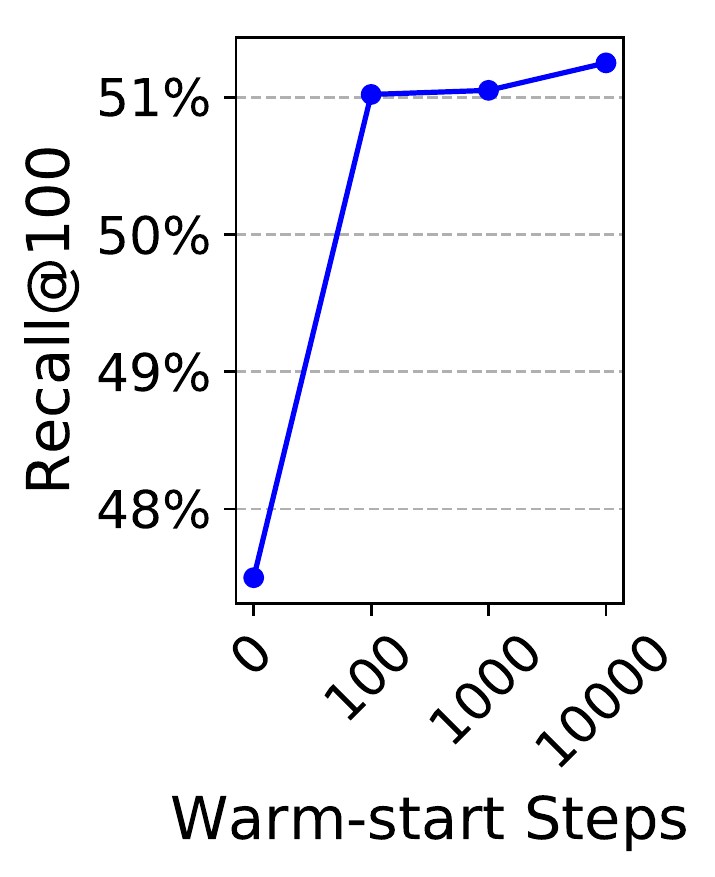}
        \vspace{-4mm}
        \caption{}
        \label{fig:warmstart}
    \end{subfigure}
    \begin{subfigure}[b]{0.3\textwidth}
        \centering
        \includegraphics[width=\textwidth]{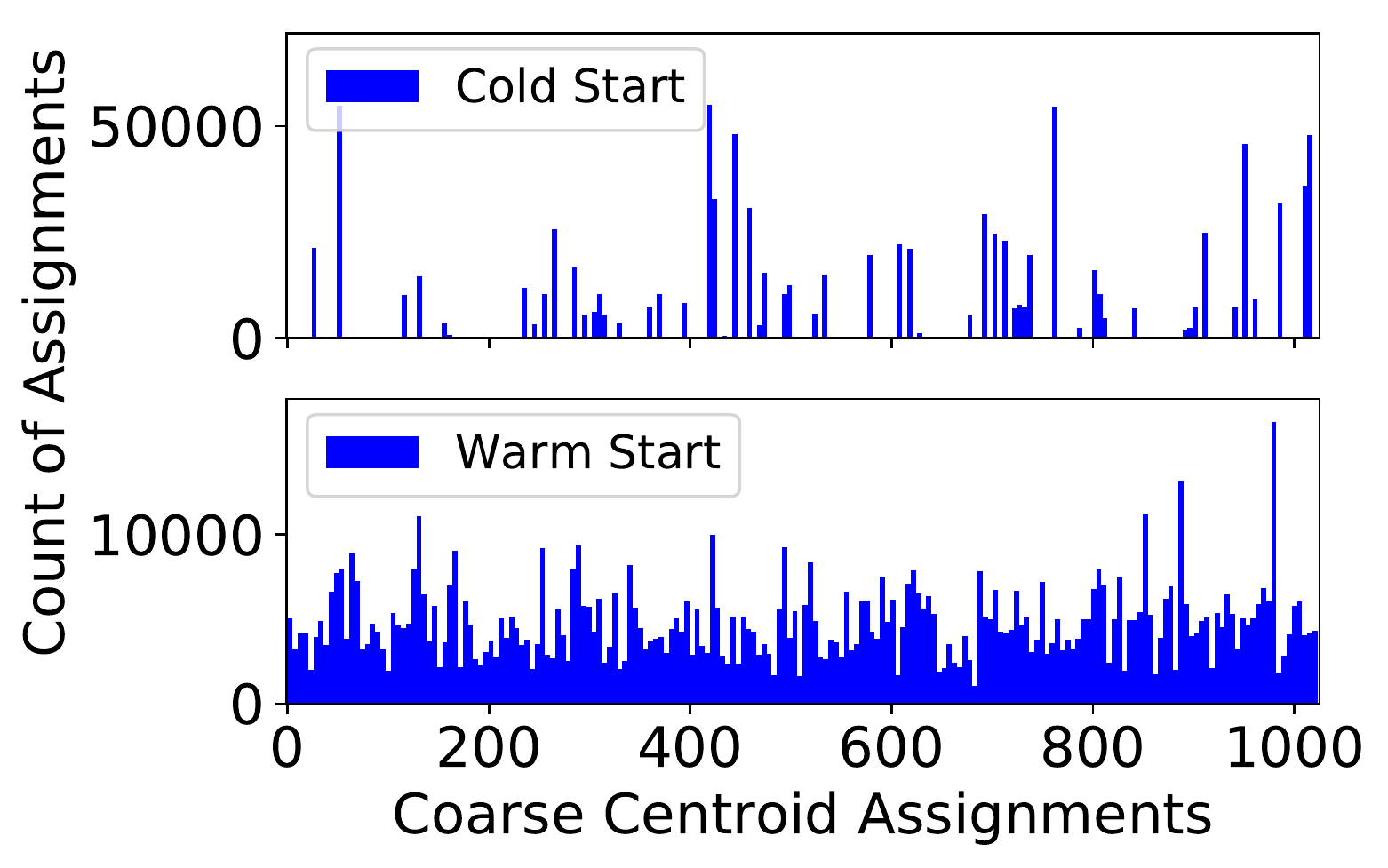}
        \vspace{-4mm}
        \caption{}
        \label{fig:histogram}
    \end{subfigure}
    \caption{Effects of warm start centroids. (a) recall@100 within different warm-start steps, (b) histogram of coarse centroid assignments with cold start (upper) and warm start (lower).}
    \label{fig:distortion}
    \vspace{-2mm}
\end{figure}

\subsection{Effects of Warm Start Centroids}
\label{sec:warm_start}
Figure~\ref{fig:warmstart} illustrates the effect of warm-start centroids, where we can observe that the retrieval quality significantly improves with warm started centroids over cold started centroids, but only slightly improves with more warm-up steps. Note that the warm-up steps $0$ corresponds to cold start where the centroids are actually randomly initialized, which results in sparse utilization of both coarse centroids and PQ centroids. In a typical run as shown in Figure~\ref{fig:histogram}, the coarse centroid assignments with warm start distribute more uniformly than those with cold start. In more detail, only $67$ out of $1024$ cold start centroids are actually used by coarse quantization, which spikes to $1004$ out of $1024$ for warm start centroids.

\subsection{Computational Cost}
\label{sec:computational_cost}
Extensive experiments show that the training time of the proposed method slightly increases by around $1\%$. And we do not observe any significant memory increasing since the coarse centroids and PQ centroids are negligible compared to the other part of retrieval model. As for the indexing time, the proposed method only needs to save items' coarse codes and PQ codes into index file. Thus, for 1 million 512-dimensional embeddings, {\pqm} only needs 5 seconds of indexing time compared to 641 seconds of Faiss, 101 seconds of ScaNN, 93 seconds of Annoy, and 4.6 seconds of LSH.

\begin{figure}[t]
    \centering
    \begin{subfigure}[b]{0.24\columnwidth}
        \centering
        \includegraphics[width=\textwidth]{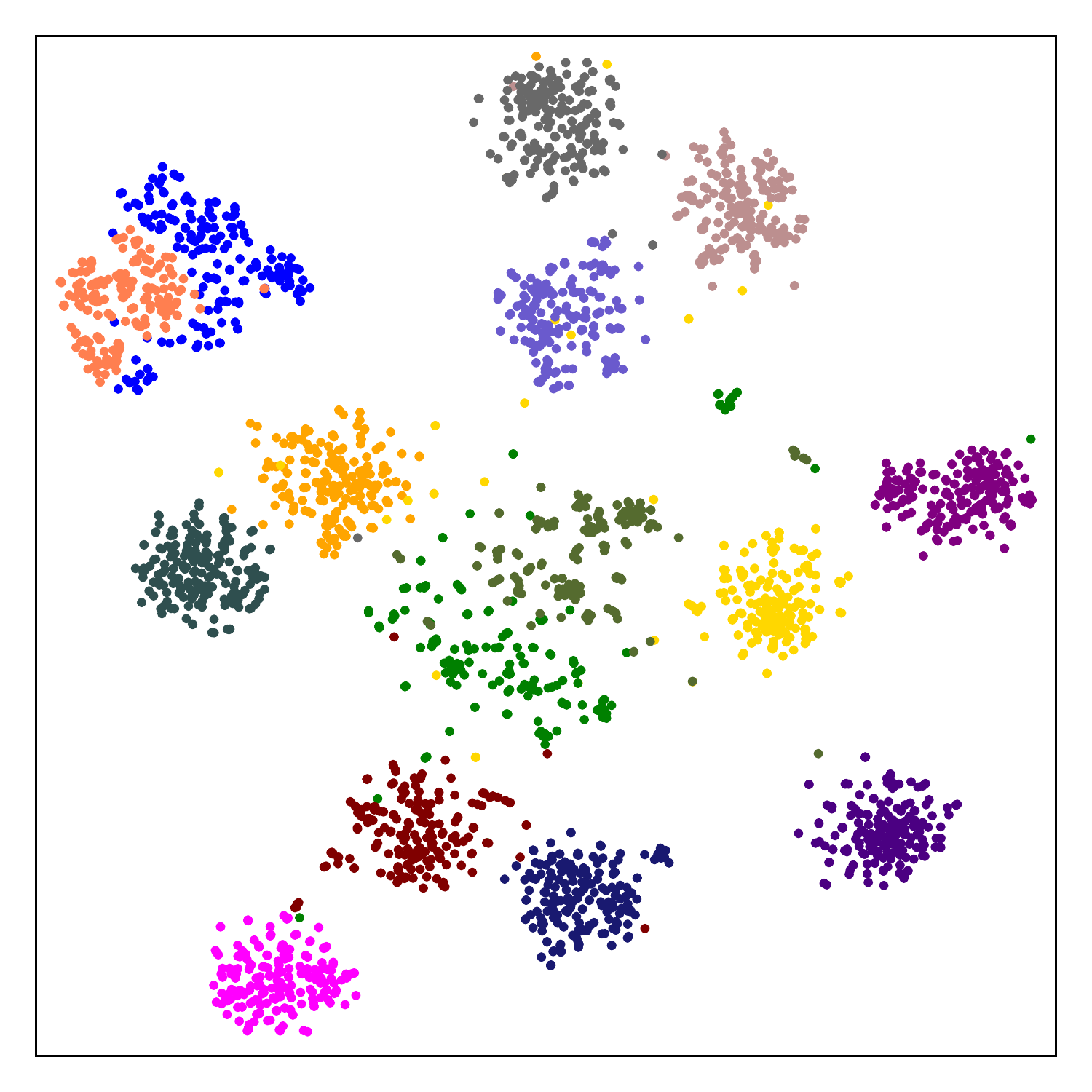}
        \caption{Raw}
        \label{fig:raw}
    \end{subfigure}
    \begin{subfigure}[b]{0.24\columnwidth}
        \centering
        \includegraphics[width=\textwidth]{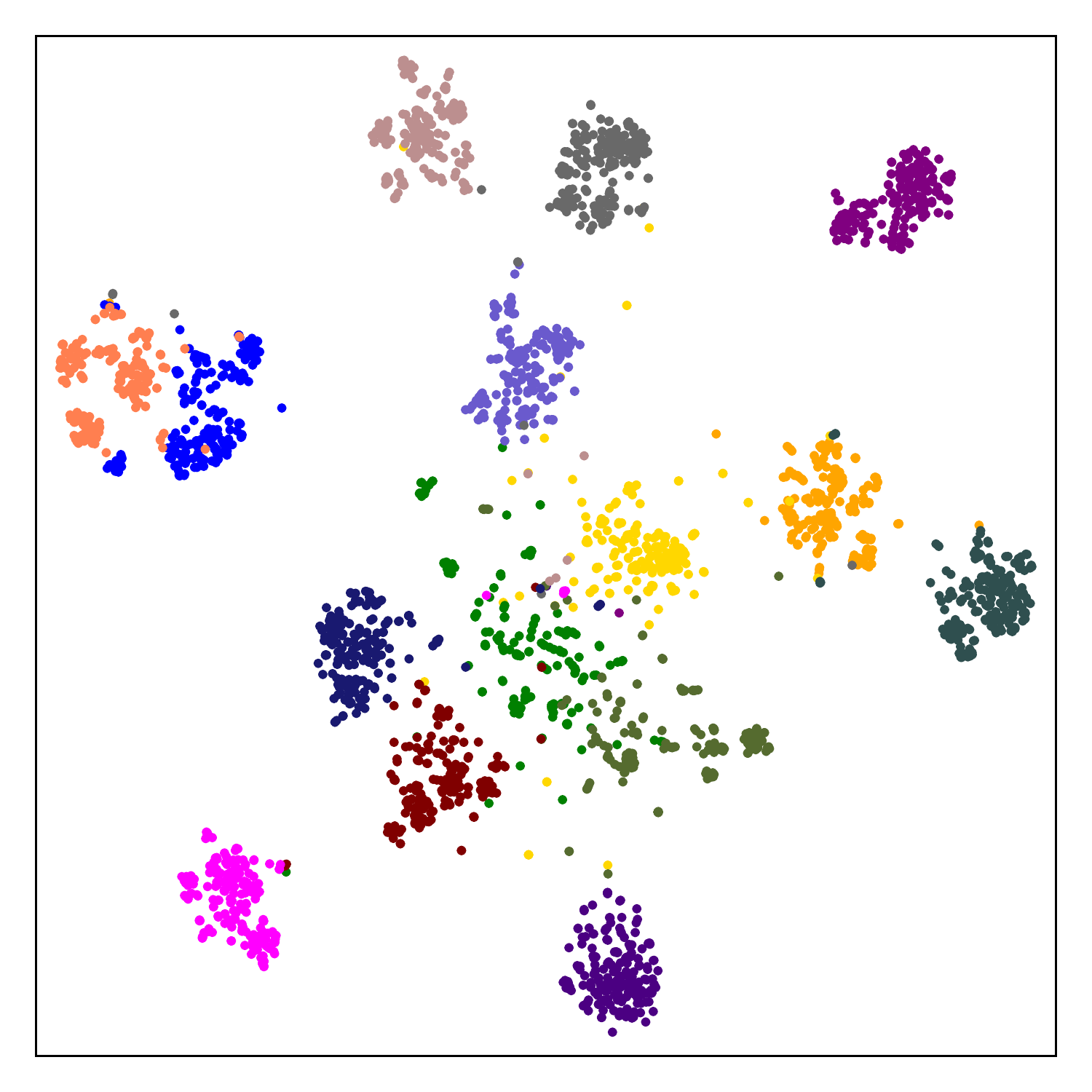}
        \caption{Faiss D64}
        \label{fig:faiss_d64}
    \end{subfigure}  
    \begin{subfigure}[b]{0.24\columnwidth}
        \centering
        \includegraphics[width=\textwidth]{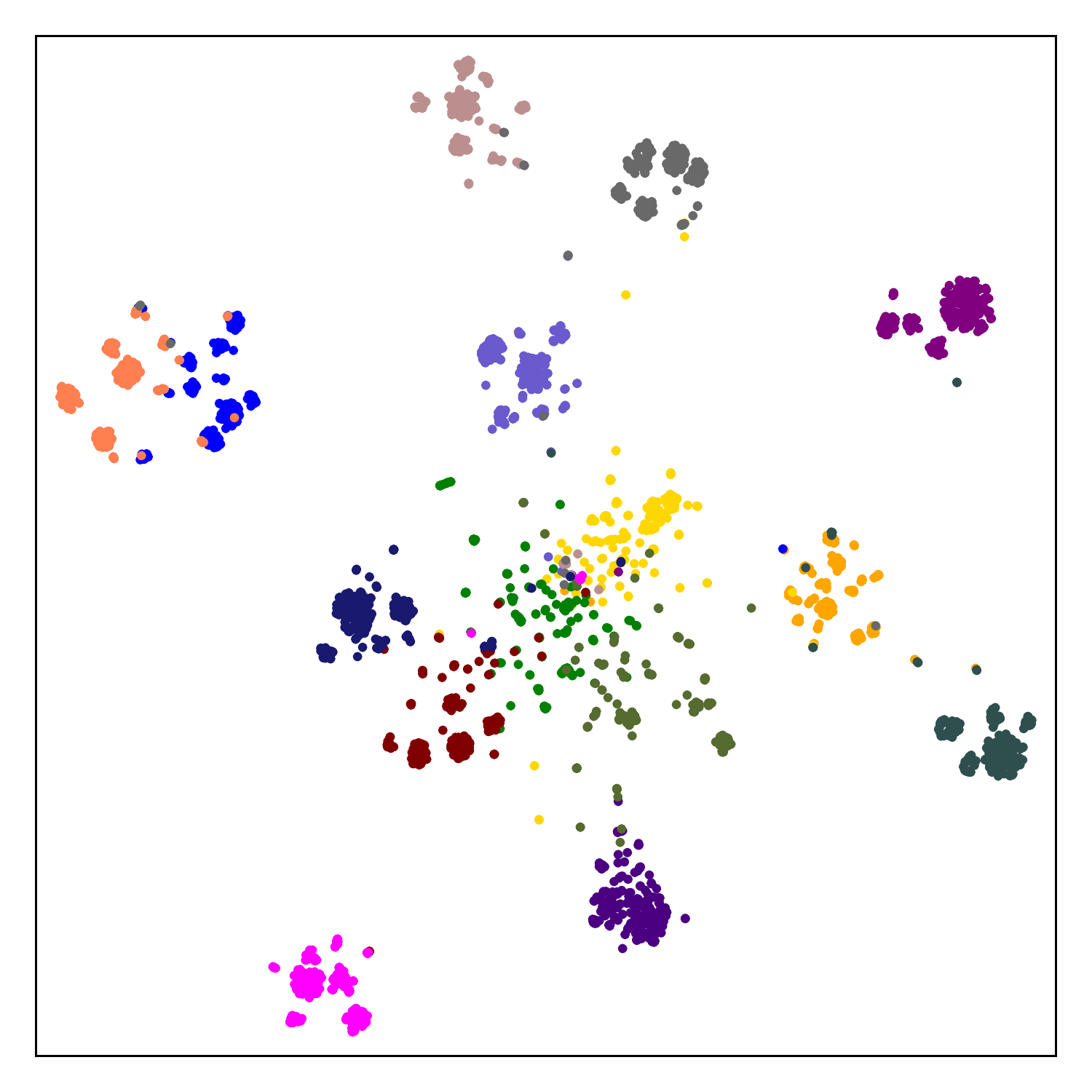}
        \caption{Faiss D32}
        \label{fig:faiss_d32}
    \end{subfigure}
    \begin{subfigure}[b]{0.24\columnwidth}
        \centering
        \includegraphics[width=\textwidth]{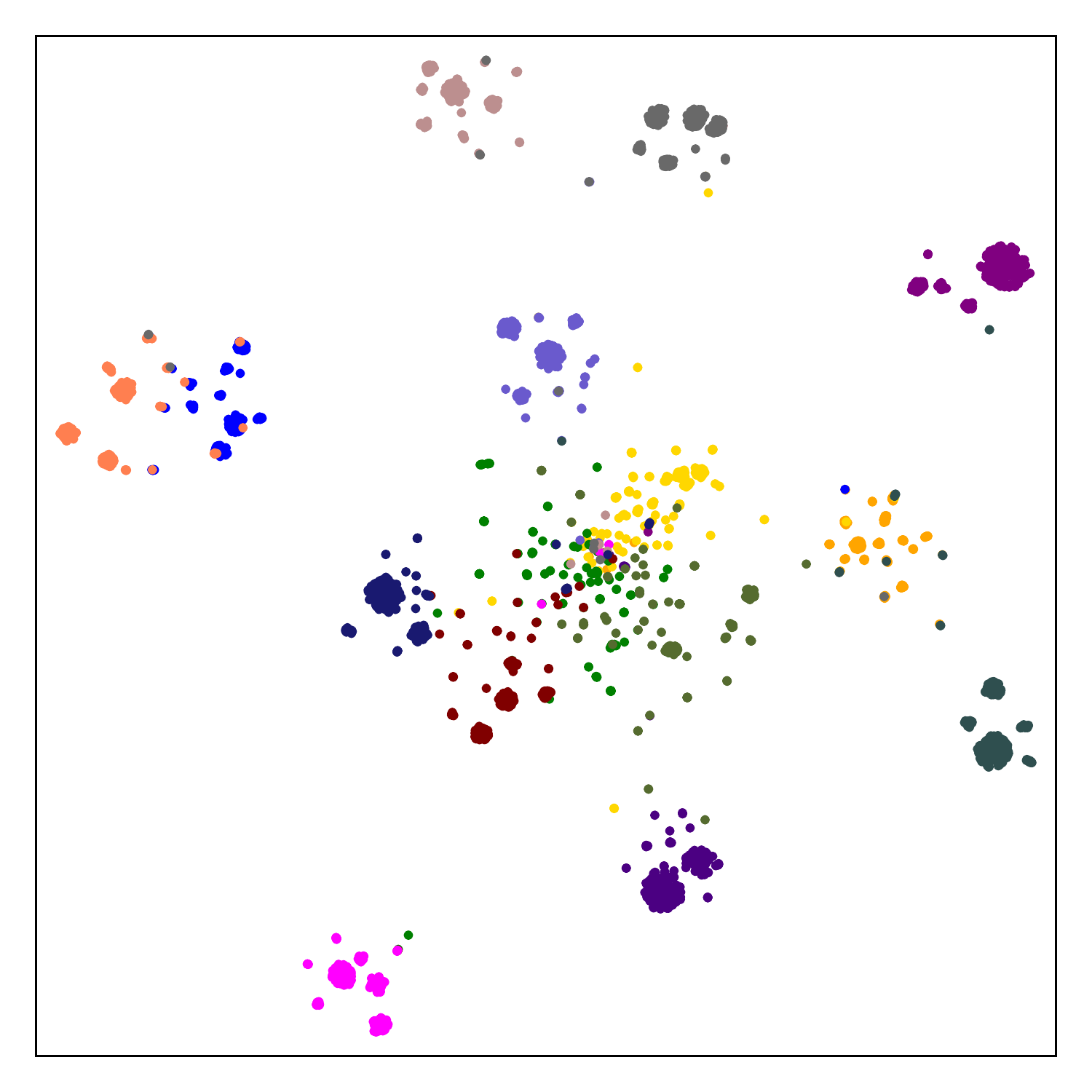}
        \caption{Faiss D16}
        \label{fig:faiss_d16}
    \end{subfigure}
     \begin{subfigure}[b]{0.24\columnwidth}
        \centering
        \includegraphics[width=\textwidth]{figures/tsne/org.pdf}
        \caption{Raw}
        \label{fig:raw2}
    \end{subfigure}
    \begin{subfigure}[b]{0.24\columnwidth}
        \centering
        \includegraphics[width=\textwidth]{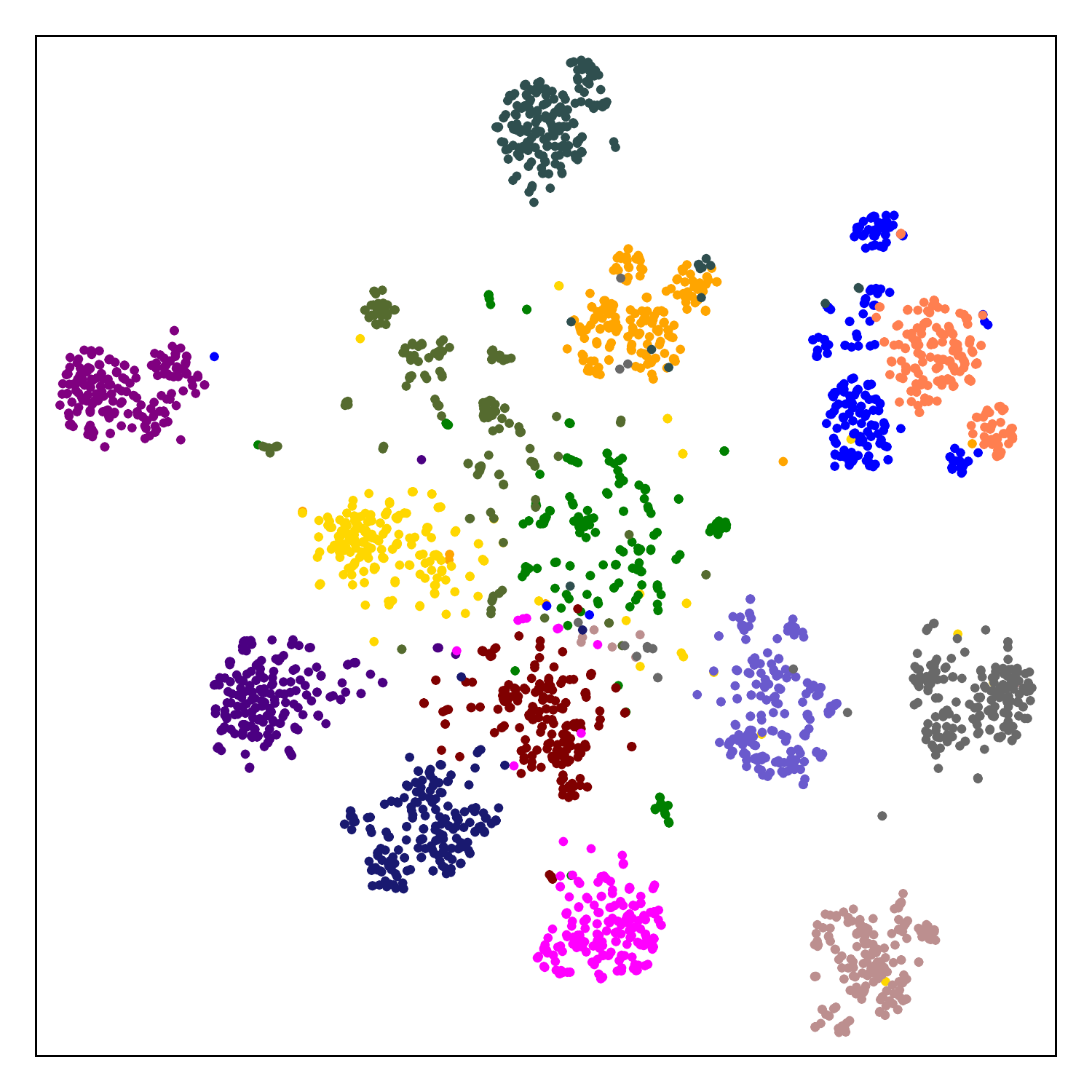}
        \caption{Poeem D64}
        \label{fig:poeem_d64}
    \end{subfigure}
    \begin{subfigure}[b]{0.24\columnwidth}
        \centering
        \includegraphics[width=\textwidth]{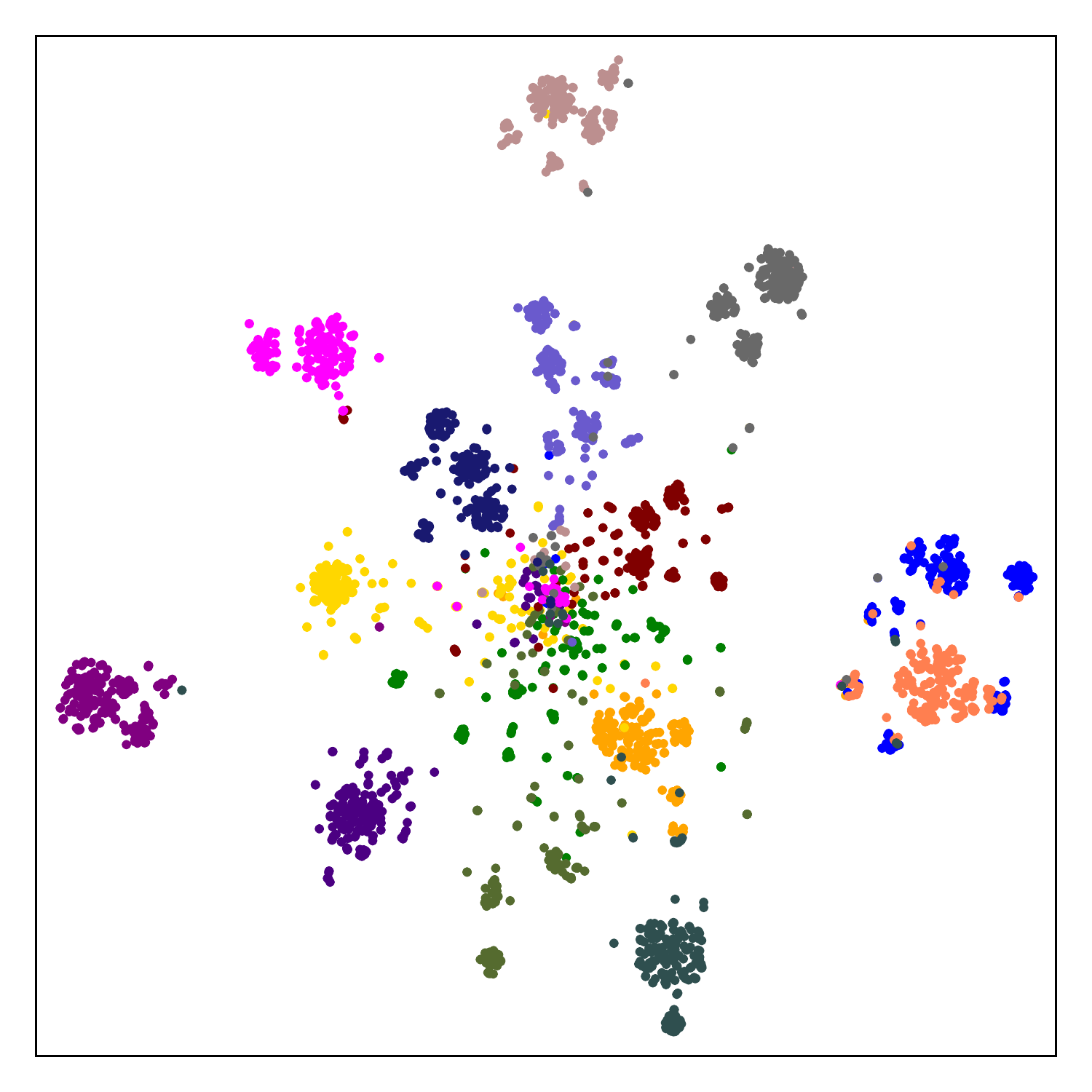}
        \caption{Poeem D32}
        \label{fig:poeem_d32}
    \end{subfigure}   
    \begin{subfigure}[b]{0.24\columnwidth}
        \centering
        \includegraphics[width=\textwidth]{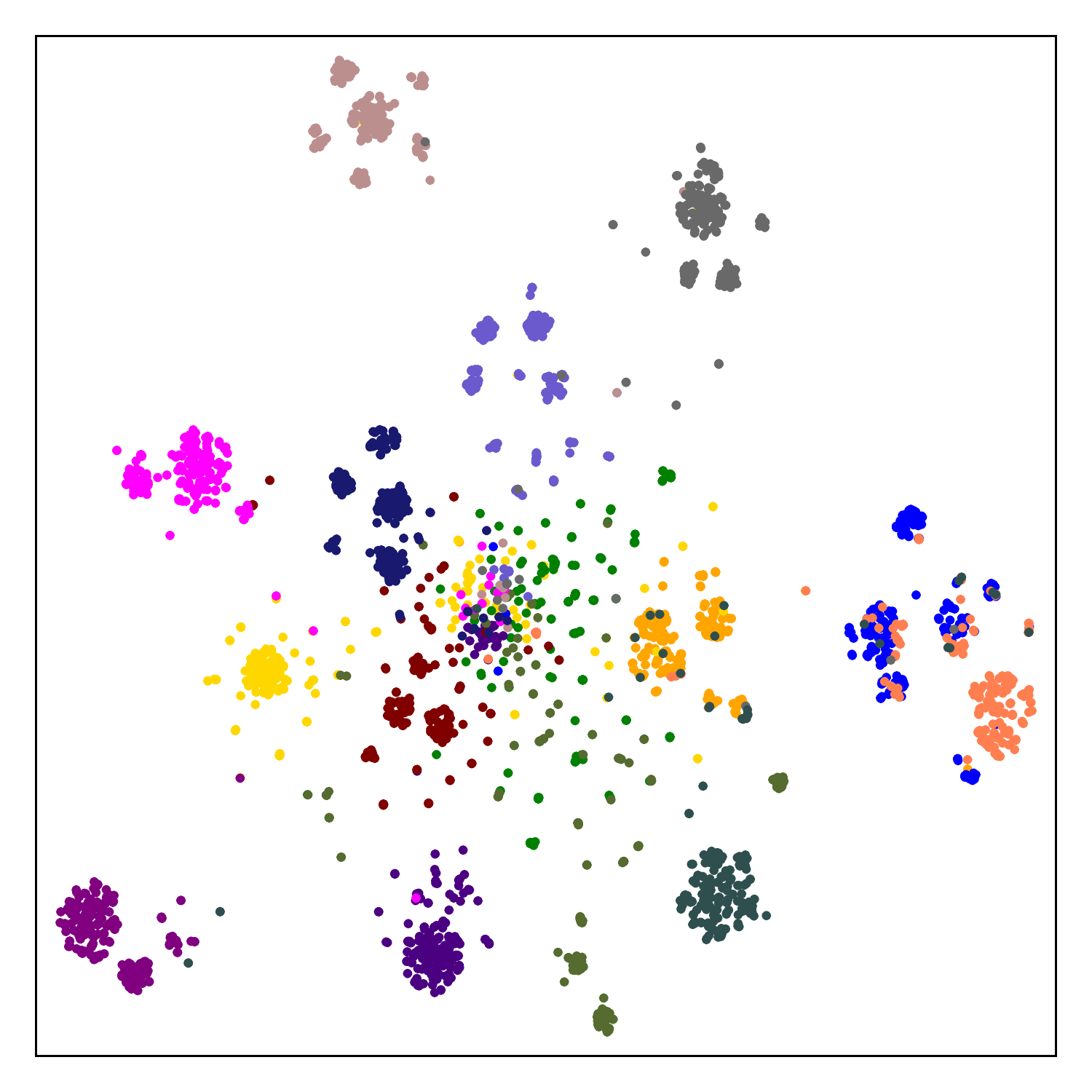}
        \caption{Poeem D16}
        \label{fig:poeem_d16}
    \end{subfigure}
    % \vspace{-0.2in}
    \caption{t-SNE visualizations of Faiss and {\pqm} item embeddings with varying parameter $D$.
    % (a), item embeddings after Faiss indexing (b$\sim$d) and {\pqm} indexing (e$\sim$g) (with same J=1024, K=256, different D=64, 32, and 16 respectively).
    }
    \label{fig:t_sne}
    \vspace{-2mm}
\end{figure}

\subsection{Ablation Study}
\label{sec:ablation}
To have an intuitive understanding of how {\pqm} works, Figure~\ref{fig:t_sne} shows 2-D t-SNE~\cite{tsne2008} visualizations on randomly chosen items from top 15 popular categories in our private dataset. Figures~\ref{fig:raw} and~\ref{fig:raw2} show the same figure of raw item embedding distribution from the two-tower retrieval model, which serves as the best scenarios since there is no quantization distortion. Figures~\ref{fig:faiss_d64} to~\ref{fig:faiss_d16} illustrate the progressive quantization distortion effect with decreasing parameter $D$ for Faiss, and Figures~\ref{fig:poeem_d64} to~\ref{fig:poeem_d16} illustrate that for {\pqm}.

For both Faiss and {\pqm}, we can observe that product quantization has the effect of ``shrinking'' and ``collapsing'' those normally distributed clusters, with the progress that the well distributed cluster is first divided into sub-clusters, which then further shrink into much smaller ones.
This effect makes sense if we consider that the product quantization forces those embeddings to share the same set of subvector centroids. Thus, those nearby embeddings are ``pulled'' closer if they are assigned to the same subvector centroid.

With the above observation, we can now see that the proposed method {\pqm} improves on the baseline Faiss, by slowing down the progress of ``shrinking'' and ``collapsing'' those clusters. While the parameter $D$ is decreasing, {\pqm} maintains the cluster's normally distributed shape much better than Faiss, especially for those outskirt clusters where the comparison is clear.

% From these visualizations, we can make the following observations: 1) items are clustered in an intuitive and explicit way, in which items are closer to those of the same category and vise versa. 2) as parameter D decreases, both {\pqm} and Faiss methods cluster the item embeddings more sparsely, which means the increasing of precision loss on raw embedding. As expected, Figure~\ref{fig:compare_D} shows that the recall accuracy is positive correlated to parameter D. 
% % what coincides the degree of distortion between raw item embeddings and embeddings after indexing in Figure~\ref{fig:t_sne}. 
% % Both after Faiss indexing and {\pqm} indexing, the distortion, which reflects intuitively in the plots is that items distributions are less like the raw items distribution, increases with the decreasing of parameter D, and results in the worse recall accuracy. 
% 3) comparing the subfigures of Faiss and {\pqm}, clusters in Faiss are way more smaller than {\pqm}. As we just discussed, the sparsity of clusters reflect how well a quantization method performs. Thus, compared with Faiss, {\pqm} lose less precision on raw embedding thanks to its joint training mechanism.
% % with respect to raw item distributions. The distortion illustrates why {\pqm} achieves better recall accuracy.

% \begin{figure}
%     \centering
%     % \includegraphics{}
%     \framebox(200,150){} 
%     \caption{T-sne visualizations of item embeddings before and after offline indexing, compared with online indexing.}
%     \label{fig:ablation}
% \end{figure}
\section{Conclusion}
\label{sec:conclusion}
In this paper, we have proposed a novel method called {\pqm} to learn embedding indexes jointly with any deep retrieval models. We introduce an end-to-end trainable indexing layer composed of space rotation, coarse quantization and product quantization operations. Experimental results show that the proposed method significantly improves retrieval metrics over traditional offline indexing methods, and reduces the index building time from hours to seconds.

% Thus, the proposed method, as a plugin layer, can be easily used in all embedding based retrieval models, to replace the offline indexing steps.

\bibliographystyle{ACM-Reference-Format}
\balance

\bibliography{references}

\end{document}